\documentclass{article}
\usepackage{spconf,amsmath,graphicx,hyperref}
\usepackage{amsmath,amsfonts}
\usepackage{algorithmic}
\usepackage{algorithm}
\usepackage{array}
\usepackage{textcomp}
\usepackage{stfloats}
\usepackage{lettrine}

\usepackage{url}
\usepackage{subcaption}
\usepackage{verbatim}
\usepackage{graphicx}
\usepackage{cite,flushend}
\usepackage{algorithm,algorithmic,amsmath,amssymb,amsthm,bbm,cite,color,graphicx,microtype,url}
\usepackage[USenglish]{babel}
\usepackage{hyperref}
\usepackage[utf8]{inputenc}
\usepackage[T1]{fontenc}

\title{Joint Active RIS Configuration and User Power Control for Localization:\\ A Neuroevolution-Based Approach}
\name{George Stamatelis, Hui Chen, Henk Wymeersch, and George C. Alexandropoulos 
\thanks{G. Stamatelis and G. C. Alexandropoulos are with the Department of Informatics and Telecommunications, National and Kapodistrian University of Athens, Greece (e-mails: \{georgestamat, alexandg\}@di.uoa.gr). H. Chen and H. Wymeersch are with Chalmers University of Technology, Sweden (e-mails: \{hui.chen, henkw\}@chalmers.se). The work has been supported by the SNS JU project 6G-DISAC under EU's Horizon Europe research and innovation programme under the Grant Agreement number 101139130. The work of the first author has been also supported by the Hellenic Foundation for Research and Innovation (HFRI) under the 5th Call for HFRI PhD Fellowships (Fellowship Number: 21080).}
}
\address{}

\begin{document}
\maketitle
\begin{abstract}
This paper studies user localization aided by a Reconfigurable Intelligent Surface (RIS). A feedback link from the Base Station (BS) to the user is adopted to enable dynamic power control of the user pilot transmissions in the uplink. 
A novel multi-agent algorithm for the joint control of the RIS phase configuration and the user transmit power is presented, which is based on a hybrid approach integrating NeuroEvolution (NE) and supervised learning. The proposed scheme requires only single-bit feedback messages for the uplink power control, supports RIS elements with discrete responses, and is numerically shown to outperform fingerprinting, deep reinforcement learning baselines and backpropagation-based position estimators.
\end{abstract}
\section{Introduction}
Reconfigurable Intelligent Surfaces (RISs) are emerging as a powerful technology for enhancing the performance of localization schemes~\cite{GAHenkLocRISPaper,GAHenkLocPaper1,ris_passive_loc_ref1,fingerprint,GAHenkHui}. However, available RIS hardware solutions~\cite{RIS_SRE} offer only discrete, or even binary, phase shifts per response-tunable unit element, making the overall RIS configuration an NP-hard problem. This has motivated the adoption of machine learning approximations, including Deep Reinforcement Learning (DRL)~\cite{AlexandroPervasive}, contextual bandits~\cite{stylianop-MAB-ICC}, and NeuroEvolution (NE)~\cite{mbacnnJournal}, with the latter being especially attractive due to its gradient-free nature being capable to handle non-differentiable objectives~\cite{salimans2017evolution}.
 
Consider a User Equipment (UE) transmitting pilot signals via the assistance of an RIS to a Base Station (BS) for the purpose of its localization. Recently, active sensing, which adaptively controls the RIS phase configuration for localization during pilot transmission, has been shown to outperform static configurations. Data-driven Neural Networks (NNs) have been presented in~\cite{activeSensingLocconf,activeSensingLocJournal}, where a  Long Short-Term Memory (LSTM) network architecture~\cite{LSTM} maintains a hidden state of past information to actively select the next RIS phase profile. After transmission, this state is passed to a second NN that estimates the final UE position. The entire system has been trained end-to-end with backpropagation, on the supervised Mean Squared Error (MSE) loss. However, the approaches in~\cite{activeSensingLocconf,activeSensingLocJournal} are limited to idealized RIS models, neglecting practical hardware limitations and impairments~\cite{RIS_SRE,11078147}.

In this paper, besides extending~\cite{activeSensingLocconf,activeSensingLocJournal} to RIS structures with reconfigurable elements of quantized responses, which makes the overall design problem for localization non-differentiable and unsuitable for standard backpropagation, we also devise UE power control. While most works optimize RISs with fixed UE power, adaptive power allocation can balance localization accuracy, privacy, and energy efficiency~\cite{powerControllPrivacy,eaht}, which constitute key performance indicators for the upcoming $6$-th Generation (6G) of wireless networks. All in all, the paper's contribution is two-fold. We formulate the problem of joint RIS phase profile selection and UE pilot transmit power control for localization, considering RISs with elements of discrete responses, which is solved via a novel Multi-Agent (MA) scheme with minimal control signaling overhead, combining NE and supervised learning. Our numerical investigations reveal the superiority of our approach over learning-based and traditional localization methods that ignore the power selection aspect.
In fact, our  method can almost match the performance of similar schemes with control signaling that results in much larger feedback costs.

\textbf{Notation:}
Lower case bold letters are used to represent vectors and upper case bold letters are reserved for matrices. $\boldsymbol{w}_x$ denotes the trainable weights of an NN $\mathcal{X}$, stacked in a vector, $\mathbb{E}[\cdot]$ denotes expectation, whereas $\Re(x)$ and $\Im(x)$ represent the real and imaginary parts of a comples number $x$.
\vspace*{-0.25cm}
\section{System Model and Design Objective}\label{sec:systemmodel}
\textbf{System Model:} We consider a system comprising a single-antenna BS receiving pilot symbols from a single-antenna UE positioned in an unknown location $\boldsymbol{p}\in\mathcal{P}_{\rm UE}\subset\mathbb{R}^3$, with the intention to obtain an accurate estimate for $\boldsymbol{p}$, denoted henceforth as $\hat{\boldsymbol{p}}$. This pilot communication is assisted by an RIS whose controller is managed by the BS via a dedicated error-free control channel. Let the RIS consist of $N_{\rm RIS}$ response-tunable elements, which, according to the vast majority of the currently available hardware implementations~\cite{RIS_SRE}, contribute an effective discrete phase shift in the impinging signal. We denote this phase shift for each $i$-th ($i=1,\ldots,N_{\rm ris}$) RIS element as $\theta_i$, with all belonging to a finite set $\Theta$. The static positions $\mathbf{p}_{\rm bs}$ and $\mathbf{p}_{\rm ris}$ of the BS and RIS, respectively, are assumed to be known to the localization framework presented in this paper. 

Block fading is assumed according to which the channels remain constant within each coherence time frame and change independently between each frame and its following one. When there is a localization request, the UE transmits a sequence of $T$ pilot symbols to the BS over an equal number of consecutive time frames. At each $t$-th ($t=1,\ldots,T$) frame, both the UE's transmission power $P(t)$ and the RIS phase profile, denoted as $\boldsymbol{\Phi}(t)\triangleq\text{diag}(\boldsymbol{\phi}(t))$ with:   
\begin{equation}
    \boldsymbol{\phi}(t)\triangleq[e^{j\pi\theta_i(t)},\ldots,e^{j\pi\theta_{N_{\rm ris}}(t)}],
\end{equation}
constitute free parameters that we intend to hereinafter optimize for our localization objective.

Let $x(t)\in\mathbb{C}$ denote the unit-power pilot symbol transmitted at the $t$-th time frame. The baseband received signal at the BS side during this frame is mathematically modeled as:
\begin{equation}
    \label{eq:received_signal}
    y(t)=P(t) \left(h_d+\mathbf{h}_{\rm{bs},\rm{ris}} \boldsymbol{\Phi}(t) \mathbf{h}_{\rm{ris},\rm{ue}}\right)x(t)+n_t,
\end{equation}
where $h_d\in\mathbb{C}$ is the gain of the direct BS-UE channel, whereas $\mathbf{h}_{\rm{bs},\rm{ris}}\in\mathbb{C}^{1\times N_{\rm ris}}$ and $\mathbf{h}_{\rm{ris},\rm{ue}}\in\mathbb{C}^{N_{\rm ris}\times1}$ represent the respective gains of the BS-RIS and RIS-UE channel matrices, and $n(t)\in \mathbb{C}$ is the additive white Gaussian noise, whose variance can be reliably estimated, thus, assumed known. Upon collecting $y(t)$, the BS decides on $\boldsymbol{\Phi}(t+1)$ and on the value of a control variable $b(t)$ to be fed back to the UE instructing it to refine its transmission power $P(t+1)$. More specifically, when the received observations are informative enough for accurate localization, the BS may request a lower uplink transmission power, thus,  enabling power savings at the UE side. On the other hand, when the received signals are too noisy to infer $\boldsymbol{p}$ satisfactorily, the BS requests higher power levels for the future pilot transmissions. To facilitate correct detection of the lightweight feedback messages at the UE side, we assume, in this paper, that $b(t)$'s are single-bit messages (e.g., ``0'' implies power reduction and ``1'' power boosting), leaving the UE to decide the exact power value within $[0, P_{\max}]$. Multi-bit messages, indicating explicit power levels at the cost, of course, of a slightly larger feedback overhead, will be considered in the journal version of this work.

\textbf{Problem Formulation:} We focus on the active sensing paradigm~\cite{activeSensingLocJournal} where, at each time instance $t$, the BS decides on the next frame's  RIS phase configuration, $\boldsymbol{\Phi}(t+1)$, leading to the most favorable observations for the UE localization objective. In addition, it also decides the transmit power level $P(t+1)$ which is acknowledged to the UE via the $b(t)$ transmission. This mode of operation implies that the RIS phase profile and the UE transmit power levels at each $(t+1)$-th time frame depend on all past observations. Let us define this dependency through a function $g(\cdot)$, i.e., it holds $\forall t<T$:
\begin{equation}
    \label{eq:policyDef}
    \left\{P(t+1),\boldsymbol{\Phi}(t+1)\right\}\triangleq g\left(y(1),\ldots,y(T)\right).
\end{equation}
The initializations $P(0)$ and $\boldsymbol{\Phi}(0)$ can be set to arbitrary values (e.g., $P(0)=P_{\rm max}$ and $\boldsymbol{\Phi}(0)$ so as to illuminate a large portion of the RIS area of influence~\cite{9827873}) if there is lack of any relevant a priori information for the localization objective. 

Following the same mindset as before, the UE position estimation at the BS after the reception of the $T$ pilot symbols will be a function of these symbols (i.e., the processing result upon them). Let $f(\cdot)$ represent this function, hence: 
\begin{equation}\label{eq:decDEf}
    \hat{\boldsymbol{p}}\triangleq f\left(y(1),\ldots,y(T)\right).
\end{equation}

Let $\mathcal{G}$ denote the set of all admissible functions $g(\cdot)$ in \eqref{eq:policyDef} and $\mathcal{F}$ the set of all estimator functions $f(\cdot)$ in \eqref{eq:decDEf}. Adopting the Euclidean distance error metric, we formulate the following optimization problem for the UE localization objective:
\begin{subequations}
\begin{equation}\label{eq:objective}
 \mathcal{OP}:    \min_{g(\cdot)\in\mathcal{G},f(\cdot)\in\mathcal{F}}\, E\left[\left\|\hat{\boldsymbol{p}}-\boldsymbol{p}\right\|_2^2\right]
\end{equation}
\begin{align}
  \text{s.t.} \quad   &\theta_i(t) \in \Theta \,\,\forall i=1,\ldots,N_{\rm ris}, \forall t=1,\ldots,T, \label{eq:risconstraint}\\ 
    &P(t) \in [0,P_{\rm max}] \,\,\forall t=1,\ldots,T, \label{eq:pmaxconstr}\\ 
    &E\left[\sum_{t=1}^T P(t)\right] \leq B_P, \label{eq:ptotalconstr}
\end{align}
\end{subequations}
where $B_P$ represents a cumulative UE power budget constraint over the $T$-frame horizon. This problem is a partially observable decision-making problem, thus, it is NP-hard~\cite{Papadimitriou1987TheCO}. In \cite{active_sensing_ref1,active_sensing_ref2,active_sensing_ref3}, data-driven approaches have been used for active sensing without the UE power control inclusion. In the sequel, we parametrize $g(\cdot)$ and $f(\cdot)$ as Deep NNs (DNNs) and propose a three-step training approach for their determination.

\section{The Proposed MA Deep Learning Method}\label{multiDL}
\begin{figure}
    \centering
    \includegraphics[width=0.85\linewidth]{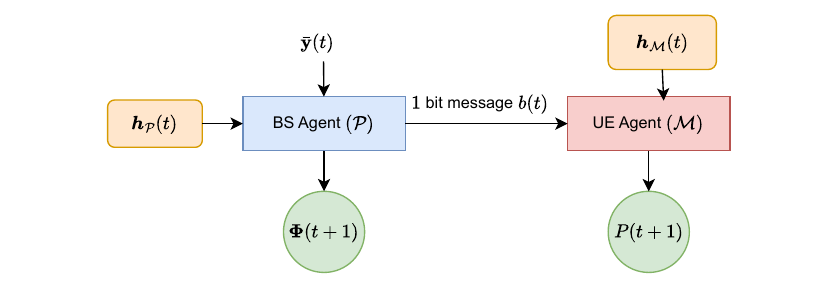}
		\caption{Graphical illustration of the proposed MA algorithm.}
    \label{fig:multiNet}
\end{figure}

In this section, we present our MA approach for the considered binary feedback messages. As illustrated in Fig.~\ref{fig:multiNet} and detailed in the sequel, the proposed approach involves two collaborating agents: one at the BS side and one at the UE. Their policies are jointly optimized to achieve the sufficient level of coordination enabling accurate localization. 

\textbf{The BS Agent:} A \textit{policy} NN, $\mathcal{P}$, and an \textit{estimator} NN, $\mathcal{E}$, are deployed, with the former controlling the RIS phase configuration as well as the BS's binary feedback to the UE, and the latter being responsible for processing the observation sequence and estimating the UE location. Both NNs include LSTMs and, for simplicity, we assume that they share the same number of hidden layers, activation functions, and layer sizes. However, they differ in their subsequent Feed-Forward (FF) stacks, which are tailored to their output requirements.

Let $\bar{\boldsymbol{y}}(t)\triangleq[\Re\{y(t)\}, \Im\{y(t)\}]$. The \textit{policy} NN takes its hidden state vector $\boldsymbol{h}_{\mathcal{P}}(t)$ and the most recent observation $\bar{\boldsymbol{y}}(t)$ as inputs to its LSTM, producing the output\footnote{Alternatively, the proposed localization approach can be implemented using the received signal strengths~\cite{active_sensing_ref1,active_sensing_ref2} (i.e., $|y(t)|^2$ $\forall t$). It will be shown in the results' section that our method performs well also with this input.} $\boldsymbol{o}^p_1(t+1)$. This output then passes through an additional NN of linear ReLU-activated layers, yielding the vector $\boldsymbol{o}^p_2(t+1) \in \mathbb{R}^{N_{\rm ris} |\Phi| +1}$. Its first $N_{\rm ris}|\Phi|$ elements are passed through an element-wise ${\rm softmax}(\cdot)$ function to define a probability distribution over the configuration of the RIS elements, while the last element is transformed using a ${\rm tanh}(\cdot)$ and a ${\rm sign}(\cdot)$ functions to produce the single-bit $b(t)$-value to be transmitted to the UE.

The \textit{estimator} NN processes the sequence $\bar{\boldsymbol{y}}(1),\ldots,\bar{\boldsymbol{y}}(T)$ by passing each of its element through the network's LSTM, producing the sequence $\boldsymbol{o}^e_1(1), \ldots, \boldsymbol{o}^e_1(T)$. This sequence is then averaged into $\tilde{\boldsymbol{o}}^e_1$, and the result is passed through a stack of linear ReLU-activated layers, resulting in an output $\hat{\boldsymbol{p}} \in \mathbb{R}^3$ that represents the estimation for the UE position.

\textbf{The UE Agent:} An additional LSTM, referred to as the \textit{power} NN, $\mathcal{M}$, is maintained at the UE side. This NN processes the most recent control value $b(t)$ along with its hidden state $\boldsymbol{h}_{\mathcal{M}}(t)$ to select the next power level $P(t+1)$. Thanks to the representational power of LSTMs, this NN can learn highly effective power control policies despite the strict communication constraint of the $1$-bit feedback. As we will demonstrate later on, the performance of this decentralized scheme can approach that of systems with much richer feedback. 

\subsection{Proposed Training Procedure}
The training of the proposed MA approach comprises three stages. First, an initial estimator NN, $\mathcal{E}_I$, is trained on a dataset of randomly generated episodes. Then, the NNs $\mathcal{P}$ and $\mathcal{M}$ are evolved to collect UE trajectories that aid $\mathcal{E}_I$'s inference capabilities. In the sequel, when the policies have been learned, a final estimator NN $\mathcal{E}$ is retrained using data collected under the learned policies of the BS and UE agents. The first and last steps involve classic supervised learning on the MSE loss function. The key step is the intermediate stage, where the NNs $\mathcal{P}$ and $\mathcal{M}$ need to be jointly optimized.

We proceed by treating the pair $(\mathcal{P}, \mathcal{M})$ as a single optimization variable within the NE framework~\cite{eaht}. Each individual in the population is represented by a concatenated parameter vector $\boldsymbol{i} \triangleq [\boldsymbol{w}_{\mathcal{P}}, \boldsymbol{w}_{\mathcal{M}}]$, and the fitness function is defined as follows: 
  \begin{equation}
        \label{eq:fitness}
    q(\boldsymbol{i}) =
    \begin{cases}
        -\hat{E}_{\mathcal{P,M}}\left[\overset{T}{ \underset{t=1}{\sum}}
 P(t)\right], \quad  \text{if } \hat{E}_{\mathcal{P,M}}\left[\overset{T}{ \underset{t=1}{\sum}}
 P(t)\right] > B_P \\
        - \hat{E}_{\mathcal{P,M},\mathcal{E}_I} \left[ \|\hat{\boldsymbol{p}} - \boldsymbol{p}\|_2 \right], \quad \text{otherwise}     \end{cases}\!\!\!\!\!\!\!,
        \end{equation}
where $\hat{E}_X$ represents sample averaging, with respect to  the random variable or set of variables $X$, over a large number $N_{\rm EP}$ of Monte Carlo episodes. In our application besides the noise $n_t$, the stochastic policies and the estimator affect the averaging.  Intuitively, this function penalizes individuals that fail to satisfy the power budget constraint $B_P$. Among those that satisfy it, individuals leading to observations of higher quality with smaller localization errors are preferred. In the NE framework, optimization is performed by first initializing a population of $L_{\rm pop}$ random individuals $\boldsymbol{i}_1,\ldots,\boldsymbol{i}_{L_{\rm pop}}$. Then, for each generation $1,\ldots,N_{\rm gen}$, each individual's parameters are split to construct the two NNs, and its fitness is calculated over multiple episodes, as follows. Individuals are shorted according to their fitness function and the best performing $\lfloor{L_{\rm pop}/4}\rfloor$ are selected for mating (crossover). During the crossover operation, the weights of two individuals are merged to construct a new individual. After that, mutation (i.e., addition of zero-mean Gaussian noise with variance $\sigma_{\rm mut}$ to each weight with probability $p_{\rm mut}$) is performed to enhance exploration. Depending on the exact choice of the NE algorithm, the percentage of individuals kept could differ, and other operations, such as gene permutations, may be included.

 \textbf{Fitness evaluation:} For each evaluation episode, a random user position $\boldsymbol{p}$ is sampled, and the hidden states of both NNs $\mathcal{P}$ and $\mathcal{M}$ are initialized. At each $t$-th time instance, the following three operations occur: \textit{i}) an observation $\bar{\boldsymbol{y}}(t)$ is sampled from~\eqref{eq:received_signal} using $P(t)$ and $\boldsymbol{\Phi}(t)$ which is provided to the policy NN $\mathcal{P}$; \textit{ii}) the latter network outputs $\boldsymbol{\Phi}(t+1)$ and an $1$-bit message $b(t)$; and \textit{iii}) the power NN $\mathcal{M}$ processes $b(t)$ to select $P(t+1)$. After $T$ steps (time frames), the observation sequence $\bar{\boldsymbol{y}}(1), \ldots, \bar{\boldsymbol{y}}(T)$ is provided to $\mathcal{E}_I$ to compute the estimation error, which determines the individual’s fitness.

\section{Numerical Evaluation}\label{sec:examples}
We have used the Cooperative Synapse NE (CoSyNE) algorithm~\cite{cosyne} with parameters $p_{\rm mut}=\sigma_{\rm mut}=0.5$, $L_{\rm pop}=50$, and $N_{\rm gen}=100$. The policy NN's LSTM was chosen to have $2$ hidden layers of $512$ units, and each of the FF branches for the selection of the RIS phase profile and $1$-one bit power parameter message had a single hidden layer of $128$ and $32$ units, respectively. The power NN’s LSTM had also $2$ hidden layers of $512$ units, followed by a linear layer of $64$ hidden units. The estimator NN was designed to have a similar LSTM followed by $2$ ReLU-activated hidden layers each of $128$ units. Both initial and final estimators were trained on $50000$ sequences. The following localization schemes have been implemented.

\begin{itemize}
\item 
Two benchmarks with power levels sampled uniformly in $[0,P_{\rm max}]$ were considered, yielding expected episodic power of $0.5\sum_{t=1}^T P_{\rm max}$. The first was a \textit{supervised} NN trained on $70000$ sequences with random RIS phase profiles using an FF model with $4$ hidden layers each with $400$ units. The other was a \textit{fingerprinting} scheme~\cite{fingerprint} where the sequence of the $T$ RIS profiles was predetermined, non-adaptive, and random. Each $1\,\mathrm{m^2}$ block in the candidate area $\mathcal{P}_{\rm UE}$ was assigned a fingerprint sequence $|y(1)|^2, \ldots, |y(T)|^2$, precomputed and stored in a database. During operation, a $5$-nearest neighbor fingerprint classifier was employed. 
\item The second step of the proposed MA scheme, which is tasked with the joint selection of $\boldsymbol{\Phi}(t)$ and $P(t)$,  was implemented via DRL, and specifically, with the popular Advantage Actor Critic (A2C) agent~\cite{A2C}. This implementation closely followed the framework in \cite{active_sensing_ref2}, adapted for localization. To enforce the power constraint, a Lagrangian reward was used similar to~\cite{SafeDRL}.
\item A scheme comprising an intelligent active sensing agent that selects continuous-valued RIS phase profiles and the UE always transmitting with $P_{\rm max}$ power has been simulated. This agent was equipped with an LSTM that decides $\boldsymbol{\Phi}(t)$ using the current observation $y(t)$ and its hidden states. The hidden states were then passed through an FF branch for the final position estimation. The structure of the LSTM layers were similar to those used in our MA method. Training was conducted using end-to-end backpropagation on the MSE loss \cite{activeSensingLocconf,activeSensingLocJournal}.
\item A single-agent variant of our MA scheme, according to which the BS transmits the exact power level it computed to the UE, has been implemented, which, however, requires substantially larger control signaling overhead.
\end{itemize}
\begin{figure*}
    \begin{subfigure}{.33\textwidth}
        \includegraphics[width=0.85\textwidth]{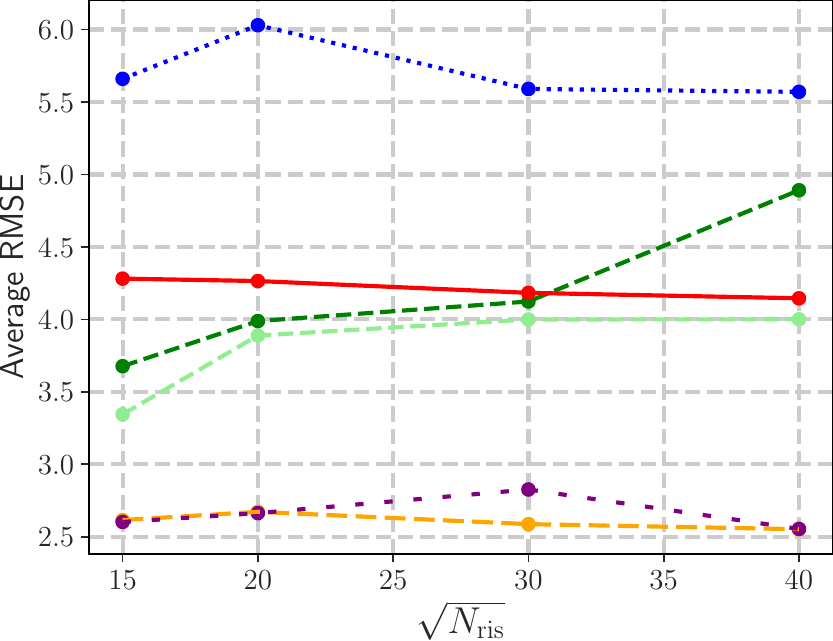}
        \caption{Stacked observations.}\label{RIS_stacked} 
    \end{subfigure} \hspace{-0.75cm}
    \begin{subfigure}{.33\textwidth}
        \includegraphics[width=0.85\textwidth]{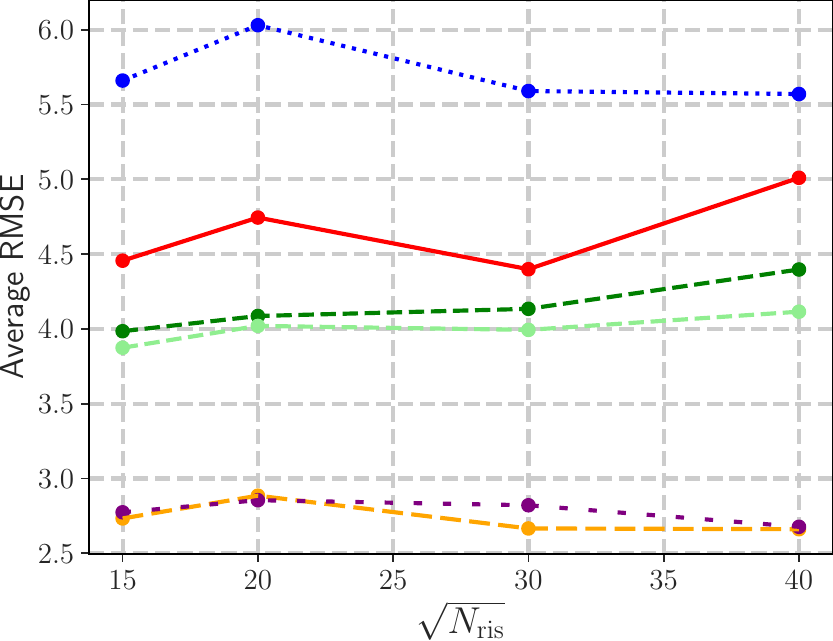}
     \caption{RSS-based observations.}\label{RIS_RSS}
    \end{subfigure} \hspace{-0.9cm}
    \begin{subfigure}{.33\textwidth}
    \centering
        \includegraphics[width=1.25\textwidth]{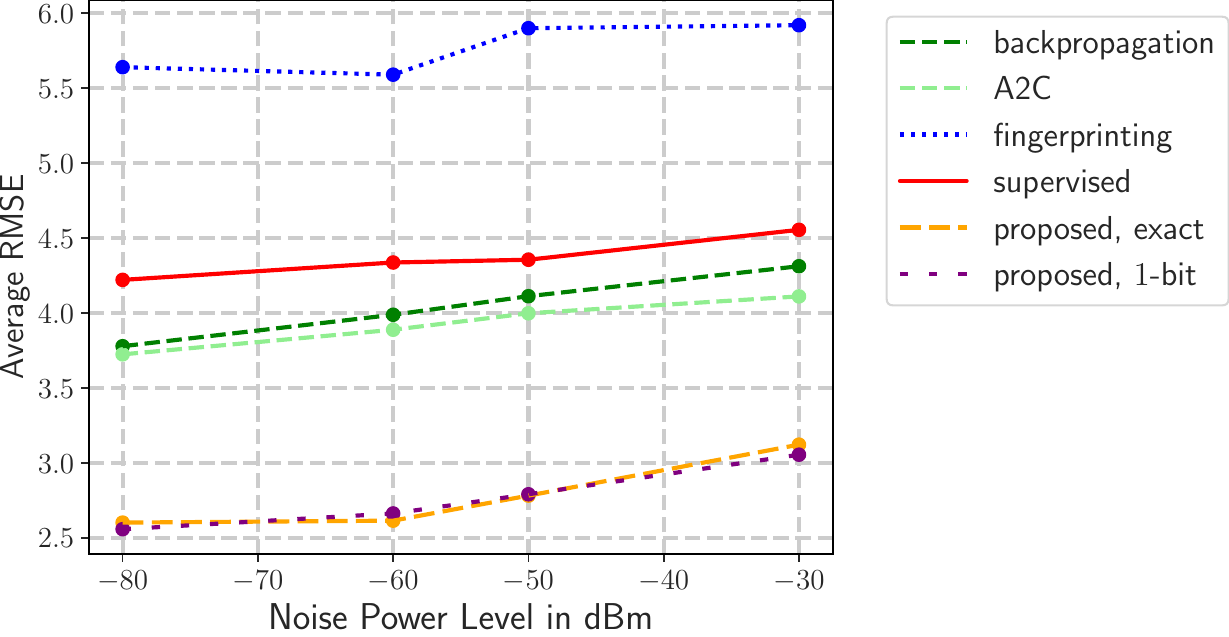}
        \caption{Stacked observations.}\label{noise}
    \end{subfigure}
    \caption{Root Mean Squared Error (RMSE) for all localization schemes considering different observation formats.}\label{fig:singleRIS}
\end{figure*}

In the conducted simulations, the UE's position was chosen as uniformly distributed inside the cubic area  $(20 \pm 15m, 20 \pm 20m,-20m)$, modeling spatial uncertainty in the $x$ and $y$ axes. Ricean fading channels, with a moderate Ricean factor of $\kappa=10$dB, and free-space pathloss were considered. The direct BS-UE channel was assumed to carry an extra attenuation of $10$dB. The BS was located at the position $(40m,-40m,10m)$ and the top-left unit element of the RIS was placed at the point $(0m,0m,0m)$. The time horizon was set to $T=10$, the maximum power  to $P_{\rm max}=30$dBm, and the power constraint to $B_{P}=0.5\sum_{t=1}^TP_{\rm max}$. 

In Fig.~\ref{RIS_stacked}, we have set the noise power level to $-60$dBm and varied the number of RIS elements from $225$ to $1600$. It can be observed that our MA scheme outperforms all benchmarks, even the one based on backpropagation that uses the full transmit power. In fact, the DRL baseline also outperforms it, indicating that training jointly the policy and the estimator NNs as in \cite{activeSensingLocJournal} is suboptimal. It is actually better to use existing supervised learning schemes to learn the estimator separately, and then fine tune the data collection policy. It is also shown that our scheme is robust to increases in the RIS size. Similar trends are demonstrated in Fig.~ \ref{RIS_RSS} for the alternative format of observations, where the DNNs are fed with the RSS values. Finally, Fig.~\ref{noise} reports RMSE performance under varying noise levels, considering an RIS with $N_{\rm RIS}=400$ elements. It can be seen that our approach achieves substantially lower errors across all noise values, highlighting its robustness. Across all simulated scenarios, our proposed MA scheme with $1$-bit control communication between the BS and UE achieves performance that is either comparable to, or only marginally inferior to, that of its  exact power value signaling variant. This indicates that the proposed power NN, $\mathcal{M}$, provides an effective mechanism for reducing control signaling overhead, while still supporting accurate localization under power constraints.

\section{Conclusion and Future Work}\label{sec:conclusion}
In this paper, we demonstrated that active sensing can effectively enhance localization  in systems with power-limited UEs. A novel hybrid scheme integrating NE and supervised learning was proposed, which was shown to outperform fingerprinting schemes, constrained DRL policies, and backpropagation-based approaches using maximum uplink UE power.
For future work, we aim to develop extensions to multi-RIS systems.

\bibliographystyle{IEEEbib}
\bibliography{strings,refs}

\begin{thebibliography}{10}

\bibitem{GAHenkLocRISPaper}
K.~Keykhosravi et~al.,
\newblock ``Leveraging {RIS}-enabled smart signal propagation for solving infeasible localization problems,''
\newblock {\em IEEE Veh. Technol. Mag.}, vol. 18, no. 2, pp. 20--28, 2023.

\bibitem{GAHenkLocPaper1}
H.~Kim et~al.,
\newblock ``{RIS}-enabled and access-point-free simultaneous radio localization and mapping,''
\newblock {\em IEEE Trans. Wireless Commun.}, vol. 23, no. 4, pp. 3344--3360, 2024.

\bibitem{ris_passive_loc_ref1}
Z.~Abu-Shaban et~al.,
\newblock ``Near-field localization with a reconfigurable intelligent surface acting as lens,''
\newblock in {\em Proc. {IEEE} {ICC}}, Montreal, Canada, 2021.

\bibitem{fingerprint}
C.~L. Nguyen et~al.,
\newblock ``Wireless fingerprinting localization in smart environments using reconfigurable intelligent surfaces,''
\newblock {\em IEEE Access}, vol. 9, pp. 135526--135541, 2021.

\bibitem{GAHenkHui}
H.~Chen et~al.,
\newblock ``{RISs} and sidelink communications in smart cities: The key to seamless localization and sensing,''
\newblock {\em IEEE Commun. Mag.}, vol. 61, no. 8, pp. 140--146, 2023.

\bibitem{RIS_SRE}
G.~C. Alexandropoulos et~al.,
\newblock ``{RIS}-enabled smart wireless environments: Deployment scenarios, network architecture, bandwidth and area of influence,''
\newblock {\em EURASIP J. Wireless Commun. Netw.}, vol. 103, pp. 1--38, 2023.

\bibitem{AlexandroPervasive}
G.~C. Alexandropoulos et~al.,
\newblock ``Pervasive machine learning for smart radio environments enabled by reconfigurable intelligent surfaces,''
\newblock {\em Proc. {IEEE}}, vol. 110, no. 9, pp. 1494--1525, 2022.

\bibitem{stylianop-MAB-ICC}
K.~Stylianopoulos et~al.,
\newblock ``Deep contextual bandits for orchestrating multi-user {MISO} systems with multiple {RISs},''
\newblock in {\em Proc. IEEE ICC}, Seoul, South Korea, 2022.

\bibitem{mbacnnJournal}
G.~Stamatelis et~al.,
\newblock ``Evolving multi-branch attention convolutional neural networks for online {RIS} configuration,''
\newblock {\em {IEEE} Trans. Cogn. Commun. Netw. to appear}, 2025.

\bibitem{salimans2017evolution}
T.~Salimans et~al.,
\newblock ``{E}volution strategies as a scalable alternative to reinforcement learning,''
\newblock {\em ar{X}iv preprint: 1703.03864}, 2017.

\bibitem{activeSensingLocconf}
Z.~Zhang et~al.,
\newblock ``Active sensing for localization with reconfigurable intelligent surface,''
\newblock in {\em Proc. {IEEE} {ICC}}, Rome, Italy, 2023.

\bibitem{activeSensingLocJournal}
Z.~Zhang et~al.,
\newblock ``Localization with reconfigurable intelligent surface: An active sensing approach,''
\newblock {\em IEEE Trans. Wireless Commun.}, vol. 23, no. 7, pp. 7698--7711, 2024.

\bibitem{LSTM}
S.~Hochreiter and J.~Schmidhuber,
\newblock ``Long short-term memory,''
\newblock {\em Neural Comput.}, vol. 9, no. 8, pp. 1735--1780, 1997.

\bibitem{11078147}
G.~C. Alexandropoulos et~al.,
\newblock ``Characterization of indoor reconfigurable intelligent surface-assisted channels at $304$ {GHz}: Experimental measurements, challenges, and future directions,''
\newblock {\em IEEE Veh. Technol. Mag.}, vol. 20, no. 3, pp. 20--29, 2025.

\bibitem{powerControllPrivacy}
O.~Arana et~al.,
\newblock ``Analysis of the effectiveness of transmission power control as a location privacy technique,''
\newblock {\em Computer Netw.}, vol. 163, pp. 106880, 2019.

\bibitem{eaht}
G.~Stamatelis et~al.,
\newblock ``Single- and multi-agent private active sensing: A deep neuroevolution approach,''
\newblock in {\em Proc {IEEE} {ICC}}, Denver, USA, 2024.

\bibitem{9827873}
G.~C. Alexandropoulos et~al.,
\newblock ``Near-field hierarchical beam management for {RIS}-enabled millimeter wave multi-antenna systems,''
\newblock in {\em Proc. IEEE SAM}, Trondheim, Norway, 2022.

\bibitem{Papadimitriou1987TheCO}
C.~H. Papadimitriou and J.~N. Tsitsiklis,
\newblock ``The complexity of {M}arkov decision processes,''
\newblock {\em Math. Oper. Res.}, vol. 12, pp. 441--450, 1987.

\bibitem{active_sensing_ref1}
F.~Sohrabi et~al.,
\newblock ``Deep active learning approach to adaptive beamforming for mmwave initial alignment,''
\newblock in {\em Proc. {IEEE} {ICASSP}}, Toronto, Canada, 2021.

\bibitem{active_sensing_ref2}
G.~Stamatelis and N.~Kalouptsidis,
\newblock ``Active hypothesis testing in unknown environments using recurrent neural networks and model free reinforcement learning,''
\newblock in {\em Proc. {EUSIPCO}}, Helsinki, Finland, 2023.

\bibitem{active_sensing_ref3}
F.~Sohrabi et~al.,
\newblock ``Active sensing for communications by learning,''
\newblock {\em {IEEE} J. Sel. Areas Commun}, vol. 40, no. 6, pp. 1780--1794, 2022.

\bibitem{cosyne}
F.~Gomez et~al.,
\newblock ``Accelerated neural evolution through cooperatively coevolved synapses,''
\newblock {\em J. Mach. Learn. Res.}, vol. 9, no. 31, pp. 937--965, 2008.

\bibitem{A2C}
V.~Mnih et~al.,
\newblock ``Asynchronous methods for deep reinforcement learning,''
\newblock in {\em Proc. {ICML}}, NY, USA, 2016.

\bibitem{SafeDRL}
A.~Ray et~al.,
\newblock ``Benchmarking safe exploration in deep reinforcement learning,''
\newblock {\em ar{X}iv preprint: 1910.01708}, 2019.

\end{thebibliography}

\end{document}